# Detecting Critical Resonances in Microwave Amplifiers through Noise Simulations

J.M. Collantes, N. Otegi, A. Anakabe, L. Mori, A. Barcenilla, J.M. Gonzalez-Perez

Dept. de Electricidad y Electrónica, University of the Basque Country (UPV/EHU), Leioa, 48940, Spain, juanmari.collantes@ehu.es

*Abstract* — The presence of critical resonances in microwave power amplifiers has a negative impact on its behavior and performance. These critical resonances are usually predicted from pole-zero stability simulations. In this paper, a different and less demanding approach for the circuit designer is proposed. It is based on performing noise simulations of the amplifier and observing the rise in the noise spectrum that happens when the system has low damping poles. Critical resonance detection is simplified since no additional probes have to be inserted in the circuit and no post-processing for pole-zero analysis is required. The technique is applied to two amplifier prototypes fabricated in microstrip hybrid technology and the results are compared with the conventional pole-zero approach.

## I. Introduction

Microwave power amplifiers are among the most critical elements in modern RF and microwave transceivers. They need to be stable, efficient and linear enough to avoid distortion of large-band complex signals with high peak to average ratios [1], [2]. A usual problem that affects the regular functioning of power amplifiers is the presence of low-frequency resonances with low damping factors. These critical resonances are associated with bias components that, combined with transistor intrinsic elements, create low-frequency parasitic loops. There are several unwanted effects associated with the presence of these critical resonances: They cause long transients and ripples in pulsed operation [3]; they hinder the linearization of amplifiers working with large instantaneous bandwidth signals [4], [5]; and, eventually, the amplifier can become unstable against circuit parameter variations, showing a spurious autonomous signal at a frequency close to the frequency of the resonance.

In the context of circuit diagnosis, it is important to have efficient tools for detecting these critical resonances. This is particularly significant in space applications, where amplifier robustness against aging or parameter changes is mandatory. Critical resonances can be predicted in simulation through pole-zero stability analyses [6]. Results of this kind of analyses provide us with the pole-zero map of the circuit, where low damping poles responsible for high resonant effects can be directly monitored. Pole-zero analysis requires the inclusion of one or several small-signal current or voltage sources and requires post-processing of the simulation results with additional software tools [7]. Unfortunately, pole-zero identification tools are not always available for the microwave circuit designer.

In this work, a different approach is proposed in order to detect the critical resonances. Instead of calculating the poles directly, the approach is based on simulating the rise in the noise spectrum that happens when the system has low damping poles. These noise bumps are often found when measuring power amplifiers [8]. They will be used here as an indication of the presence of the critical resonances associated with complex conjugate poles with very low damping. In this way, critical resonances can be detected without including any additional probes in the circuit and without using any post-processing tool for pole-zero identification. It is important to note that the technique is not a stability analysis. It only makes sense for stable poles.

The paper is organized as follows. Section II describes briefly the nature of these noise bumps, also called noisy precursors. In section III, the approach is explained and tested on an L-band amplifier built in microstrip hybrid technology in which low-frequency critical resonances are correlated to simulated noise bumps. In section III, the simulation of the noise bumps in a three-stage amplifier is correlated to the residue analysis presented in [6], in order to detect the origin of the critical resonance.

## II. Noisy Precursors

Let us consider a linear or linearized circuit having a pair of complex conjugate poles ($\sigma \pm j\omega_i$) with small negative real part. Face to a perturbation, the transient of the circuit is dominated by these low damping poles. Assuming a constant perturbation due to noise, a rise in the output noise spectrum about the frequency $\omega_i/2\pi$ can be observed. Very often, as a parameter of the circuit is modified, the critical poles end up crossing to the right half side of the complex plane and a spurious signal appears in the output spectrum. That is why these noise bumps are also called noisy precursors. A thorough explanation on the theoretical basis, modeling and simulation of the noisy precursors in autonomous microwave circuits can be found in [9].

## III. SIMULATION OF THE NOISY PRECURSORS IN AN L-BAND AMPLIFIER

With the aim of detecting low-frequency critical resonances, we have selected a prototype similar to the one presented in [10]. This is a medium power one-stage L-band amplifier built in microstrip technology with a single GaAs FET device (Fig. 1).

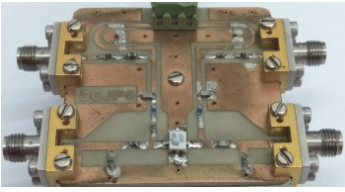

Fig. 1. Medium power one-stage L-band amplifier with a single GaAs FET.

Let us consider first the amplifier in a DC steady state with no large-signal applied. A conventional pole-zero stability analysis has been performed using ADS as electrical simulator and STAN as identification tool. Fig. 2a shows the evolution of a critical pair of complex conjugate poles, with very low abs($\sigma$) as $V_{DD}$ is varied from 5 to 8 V, with $V_{GG}$ = - 1.8 V. Frequency of the poles is about 165 MHz. The pole evolution of Fig. 2a will serve as a reference for the noise simulations. Next, an AC noise analysis is performed in ADS. Note that, we are only concerned with the detection of the critical resonance, not with the actual noise level that might be measured at the output of the amplifier. Besides, our goal is to provide the simplest procedure for the designer, actually, simpler than a pole-zero analysis. Therefore, only thermal noise of the resistive components is considered in the noise simulation and a 1 MHz noise bandwidth has been chosen. The resulting noise spectrum at the output terminal of the amplifier is depicted in Fig. 2b for the same sweep of $V_{DD}$. We can observe how the noise level exhibits a clear rise around the frequency of the critical poles. This noise bump gets higher and narrower as the $V_{DD}$ value increases, i.e. as the critical poles shift rightwards on the LHP, in consistency with Fig. 2a.

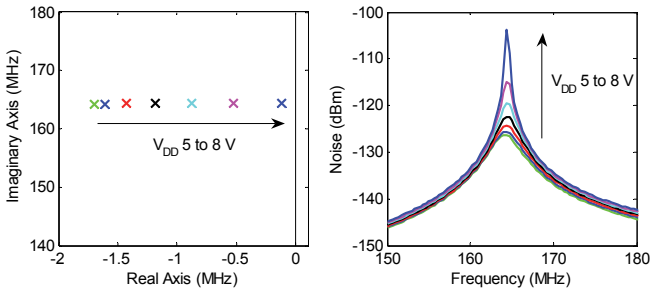

Fig. 2. a) Evolution of the critical poles with $V_{DD}$ 5 to 8 V and $V_{GG}$ = - 1.8V obtained from identification. Only positive frequencies are plotted for simplicity; b) Noise at the amplifier output resultant from AC noise analysis.

Let us analyze now the circuit driven by a large-signal input generator with input frequency $f_{in}$ and input power $P_{in}$. The linearization of this large-signal steady state is a Periodic Linear Time Variant (PLTV) system from which local stability and noise behavior can be determined. Poles of the PLTV system correspond to the Floquet exponents and repeat in frequency as $\sigma \pm j(\omega \pm k\omega_{in})$, with $k \in \mathbb{N}$. Again, to have a valid reference, we perform a pole-zero stability analysis of the circuit varying $P_{in}$ from -30 to 10 dBm, for $f_{in}$ = 1.2 GHz, $V_{DD}$ = 7 V and $V_{GG}$ = - 1.8 V. The evolution of the critical poles (crosses) in the complex plane is plotted in Fig. 3a. The pole for the same bias condition, without large signal input, from the previous AC simulation has been superimposed for comparison (square). We can observe how the low frequency poles evolve rightwards (reducing stability margin) as $P_{in}$ is increased. Actually, for Pin = 8 dBm the fabricated circuit oscillates at 170 MHz (Fig. 4), which is a little bit lower input power than the one predicted by the model.

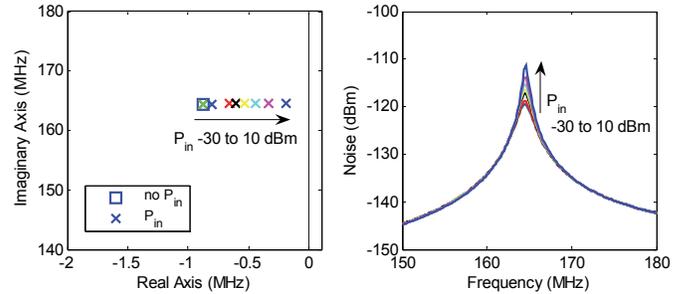

Fig. 3. a) Evolution of the critical poles with $P_{in}$ from -30 to 10 dBm obtained from identification. Only positive frequencies are plotted for simplicity; b) Noise at the amplifier output resultant from non-linear noise analysis.

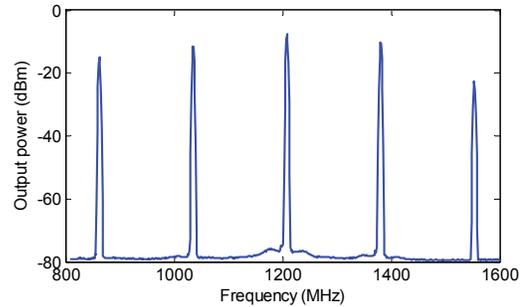

Fig. 4. Output spectrum of the L-band amplifier of Fig. 1a measured with $P_{in}$ = 8 dBm at $f_{in}$ = 1.2 GHz ($V_{GG}$ = - 1.8 V, $V_{DD}$ = 7 V). The mixing terms due to a low frequency spurious oscillation at about 170 MHz are observed.

The steady state being forced, the circuit noise spectrum can be calculated with the conversion matrix approach [9]. In this case we use the non-linear noise analysis from ADS that is based on conversion matrix. Note that, conversion matrix is also the core algorithm of the harmonic balance simulation with small-signal mode used for the pole-zero stability analysis of Fig. 3a. The critical frequencies are far from $f_{in}$. Therefore, only white noise sources are important for the noise simulation. Again, only the resistor's thermal noise is taken into account to simplify the procedure. The noise spectrum predicted at the circuit output node is given in Fig. 3b. The critical low-frequency resonance and its evolution with $P_{in}$ is well predicted again by the rise in the noise spectrum.

## IV. SIMULATION OF THE NOISY PRECURSORS IN A THREE-STAGE AMPLIFIER

The case of the three-stage power amplifier at 1.2 GHz presented in [6] is considered here. A photograph of the amplifier is shown in Fig. 5. The circuit exhibits a critical resonance at about 125 MHz versus input power [6]. A residue analysis performed at six nodes (input and output nodes of each stage) showed that the origin of the problem was in the second stage (Fig. 6a). The non-linear noise analysis described in the previous section has been performed at the same six nodes for $P_{in}$ = - 8 dBm. Note that this simulation does not require the sequential inclusion of six small-signal sources as the residue pole-zero analysis does. The noise results for the output nodes of the three stages are plotted in Fig. 6b. It is clearly noticeable that the largest noise bump is obtained at the second stage node. This is in agreement with the residue analysis shown in Fig. 6a. As a result, noise simulation at different nodes can also be used to detect the origin of the undesired dynamics.

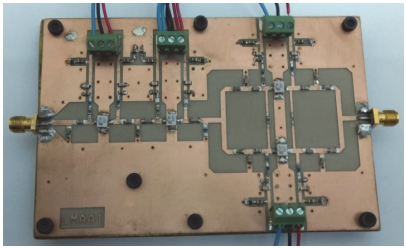

Fig. 5. Photograph of GaAs FET based L-band three-stage amplifier in microstrip technology.

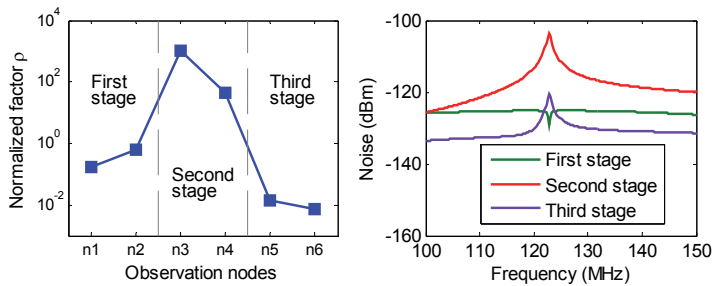

Fig. 6. a) Residue analysis of critical poles at about 125 MHz indicating that the parametric instability is at the second stage ($P_{in}$ = - 8 dBm). b) Noise results for the output nodes of the three stages ($P_{in}$ = - 8 dBm).

The noise simulation at the six nodes is almost five times faster than the simulation for stability analysis at those same six nodes. Besides it does not require the inclusion of additional probes and bypasses the need of using additional post-processing tools.

## V. DISCUSSION AND CONCLUSION

Noise simulations have been proposed here as an undemanding way to detect critical resonances in microwave amplifiers. For the circuit designer, this procedure is simpler and faster than a complete multi-node pole-zero stability analysis and prevents the use of extra tools for post-processing the simulated data.

ACKNOWLEDGEMENT

Authors wish to acknowledge the Spanish and Basque Administrations for the financial support of this work through projects TEC2015-67217-R (MINECO/FEDER) and IT1104-16 respectively.